\documentclass[twoside,10pt]{article}
\usepackage{extsizes}
\usepackage[super,sort&compress,comma]{natbib} 
\usepackage[version=3]{mhchem}
\usepackage[left=1.5cm, right=1.5cm, top=1.785cm, bottom=2.0cm]{geometry}
\usepackage{balance}
\usepackage{mathptmx}
\usepackage{graphicx} 
\usepackage{lastpage}
\usepackage[format=plain,justification=justified,singlelinecheck=false,font={stretch=1.125,small,sf},labelfont=bf,labelsep=space]{caption}
\usepackage{float}
\usepackage{fancyhdr}
\usepackage{fnpos}
\usepackage{tabularx}
\usepackage[english]{babel}

\usepackage{booktabs}
\usepackage{mhchem}
\addto{\captionsenglish}{%
  
}
\usepackage{array}
\usepackage{droidsans}
\usepackage{charter}
\usepackage[T1]{fontenc}
\usepackage[usenames,dvipsnames]{xcolor}
\usepackage{setspace}
\usepackage{hyperref}
\usepackage{multicol}

\usepackage{epstopdf}

\begin{document}
\title{{Multistate Coupled Diabatic Neural Network  potential for the quantum non-adiabatic  Photofragmentation of CH$_2^+$\footnote{~ Physical Chemistry Chemical Physics, 2026, DOI: 10.1039/D6CP01221C,
      Electronic supplementary information (ESI) available at DOI }
  }}

\author{Pablo del Mazo--Sevillano\footnote{~Departamento de Qu{\'\i}mica F{\'\i}sica, Facultad de Ciencias Qu{\'\i}micas, Universidad de Salamanca, 37008 Salamanca, Spain.  E-mail: pablomazo@usal.es}
  , Susana G\'omez--Carrasco{$^\dag$},
  Alfredo Aguado\footnote{~Departamento de Qu{\'\i}mica F{\'\i}sica Aplicada (UAM), Unidad Asociada a IFF-CSIC,
Facultad de Ciencias M\'odulo 14, Universidad Aut\'onoma de Madrid, 28049,
Madrid, Spain.} ,
and Octavio Roncero\footnote{~  Instituto de F{\'\i}sica Fundamental (IFF-CSIC), C.S.I.C., Serrano 123, 28006, Madrid,
Spain. E-mail: octavio.roncero@csic.es}
}

  \maketitle

  \section*{Abstract}
   Tracking the complex non-adiabatic transitions in far-ultraviolet photodissociation demands highly accurate
   diabatic potential energy matrices (PEMs) across numerous excited states. To address this, we introduce a fully automated
   diabatization method that leverages artificial neural networks to fit PEMs. 
   Our approach divides the PEM into a physically grounded zeroth-order diagonal term, which is then corrected by a neural network matrix to capture electronic couplings.
   By enforcing symmetry constraints on off-diagonal elements and sharing degenerate diabatic states between the $A'$ and $A''$ irreducible representations,
   the { diabatization} process becomes completely automatic.
   We validate this method using time-dependent wavepacket calculations to simulate the photodissociation of CH$_2^+$, incorporating relevant states up to $\approx 13.6$~eV.
   Finally, we compute partial cross-sections for all fragmentation channels---including total and partial fragmentation yielding
   \ce{CH+}, \ce{CH}, \ce{H2}, and \ce{H2+} diatoms---revealing a notably high cross-section for the formation of the \ce{CH} radical.
   
\begin{multicols}{2}

\section{Introduction}

Photodissociation plays a fundamental role in outer atmospheres, on Earth and exo-planets, as well as in the borders of  molecular clouds and protoplanetary disks \cite{Hemert-vanDishoeck:08,Dishoeck-Visser:15,Heays-etal:17,Hrodmarsson-vanDishoeck:23}, by controlling the abundance of small molecules. Those environments are irradiated by high-energy ultraviolet photons  producing a large variety of radicals and ions, which in turn trigger the formation of more complex molecules. Hence, it is not only crucial to know the total photodissociation rates, but also the branching ratios for the formation of photo-products in different electronic states.

The fragmentation dynamics at high energy is governed by  non-adiabatic transitions among many excited electronic states.
When these states cross, leading to conical intersections (CI), the dynamical simulations become complicated, since the non adiabatic couplings (NACs) present singularities~\cite{Yarkony:95,Yarkony-cap2-conicalbook:04}. This problem is frequently overcome by a unitary transformation to a diabatic electronic representation \cite{Smith:69,Ruedenberg-Atchity:93} where the kinetic adiabatic couplings vanish. However, this is in general not possible for all degrees of freedom \cite{Mead-Truhlar:82}, and the term quasi-diabatization is commonly used. Several diabatization procedures exist and they are classified as derivate-, property- or energy-based  methods, depending on the information they use~\cite{Koppel-cap4-conicalbook:04}. One common approach is the regularization \cite{Thiel-Koppel:99}, {\it i.e.} the elimination of the singularities appearing at conical intersections.

Block diagonalization \cite{Pacher-etal:91,Pacher-etal:93} and effective Hamiltonian \cite{Papas-etal:08} methods use reference geometries to generate a local description of diabatic Hamiltonians and are successfully applied to study photodissociation spectra. 
{
H$_3^+$ \cite{Ghosh-etal:17}, C$_6$H$_6^+$ \cite{Mukherjee-etal:21} and C$_4$N$_2$H$_4$ \cite{Hazra-etal:22} are representative examples of diabatization, using derivative-based methods, of large systems including four or five electronic states, using normal coordinates  
thus allowing the accurate description of photoabsorption spectra using quantum time dependent methods, but not fully follow the dissociation dynamics.
}
However,  in many cases there are multiple crossings, not only in the Frank-Condon region but also close to products in different rearrangement channels, that need to be considered to properly describe the branching ratios in photodissociation. The problem becomes more complex when dealing with large energy intervals, as in astrochemistry, because a large number of electronic states are needed to describe the photofragmentation rate in all the UV radiation field.

The application of machine learning techniques to produce diabatic potential energy matrices (PEM) is being found to be extremely useful as has been already proven in different systems~\cite{10.1063/1.5099106,C8CP06598E,D1CP03008F,williamsNeuralNetworkDiabatization2018}. For two-state systems fully automatic { diabatization} methods (meaning no diabatic information is used) can provide reasonable PEM from pure adiabatic information\cite{Xie-etal:18,Li-etal_SiH2p:24,Li-etal_CH2p:24}, leveraging the symmetry restrictions in the couplings of the PEM. As the number of states grows larger semi-automatic { diabatization} methods, as the family of diabatization by deep neural networks (DDNN~\cite{Shu-Truhlar:20}) , exists, where minimal diabatic information is required~\cite{shuParametricallyManagedActivation2024,delMazo25_etal_CHp}. Other approaches use machine learning methods with clustering and regression techniques to generate smooth and well ordered diabatic states with property-based diabatization \cite{Srsen-etal:24}. In this work we present a fully automatic diabatization method applicable to many electronic states over a large energy interval
{ for triatomic systems}.
 For that we first construct an initial PEM model for describing the products in different rearrangements, which are correlated by symmetry considerations allowing to impose an ordering of the diabatic states. A many--body term expressed as an artificial neural network is added to the zeroth order model to correct the short distance region and include the couplings among the diabatic states with symmetry restrictions.

The diabatization method is applied to the study of the photodissociation of CH$_2^+$ cation, whose ground state presents important Renner-Teller effects due to its $^2\Pi$ character \cite{Jensen-etal:95} and it has multiple ionic crossings in the CH$^+$/CH and H$_2$/H$_2^+$ products rearrangement channels, shown in Fig.~\ref{fig:diatomics_abinitio}, thus providing a good benchmark. This system is a missing stone in the hydrogenation chain CH$_n^+$ + H$_2$ $\rightarrow$ CH$_{n+1}^+$+H, giving rise to CH$^+_3$ which is considered the origin of hydrocarbons in the interstellar medium (ISM) \cite{Black-Dalgarno:76,Smith:92}. CH$^+$ is one of the first  molecules observed in the ISM~\cite{Douglas-Herzberg:41}, and since then has been observed in a great variety of interstellar and circumstellar environments. On the other hand, CH$_3^+$ was observed very recently in a protoplanetary disk with the James Webb Space Telescope\cite{Berne-etal:23}. Thus the missing detection of CH$_2^+$ is fundamental to corroborate the above mentioned hydrogenation chain as the formation mechanism of CH$_3^+$. Formation and destruction rates are needed to properly model the abundance of these ions. The photodissociation cross sections have been calculated for CH$^+$ \cite{Saxon-etal:80,Kirby-etal:80} and CH$^+_3$ \cite{delMazo-Sevillano-etal:24} using quantum dynamical methods, while for CH$_2^+$ only a vertical excitation model from the ground equilibrium geometry is available \cite{Theodorakopoulos-Petsalakis:91}. 

The final goal of this work is to study the non-adiabatic photofragmentation dynamics of CH$_2^+$ in a new diabatic model composed of 16 coupled electronic states, and it is organized as follows. First, the new diabatization method is presented and applied to CH$_2^+$. Next, the results on the photodissociation dynamics obtained with a quantum wave packet treatment are discussed. Finally, some conclusions are extracted.

\section{Diabatization} 
Machine learning diabatization methods often rely on some diabatic information to impose the correct order of the different states. In this work we present a fully automatic { diabatization} method based on a many body expansion of the potential energy. It is applied to produce a diabatic potential energy matrix (PEM) of 16 states of the \ce{CH2+} system, 8 in each $A'$ and $A''$ irreducible representations of the $C_s$ point group. With this number of states we properly include the Frank-Condon region of the \ce{CH2+} and possible dissociation channels up to $\approx13.6$ eV. While the diabatization method is automatic, it still requires a very detailed analysis of the electronic structure of the system. 

\subsection{Diatomic fragments}

\begin{figure}[H]
\centering
    \includegraphics[width=.8\linewidth]{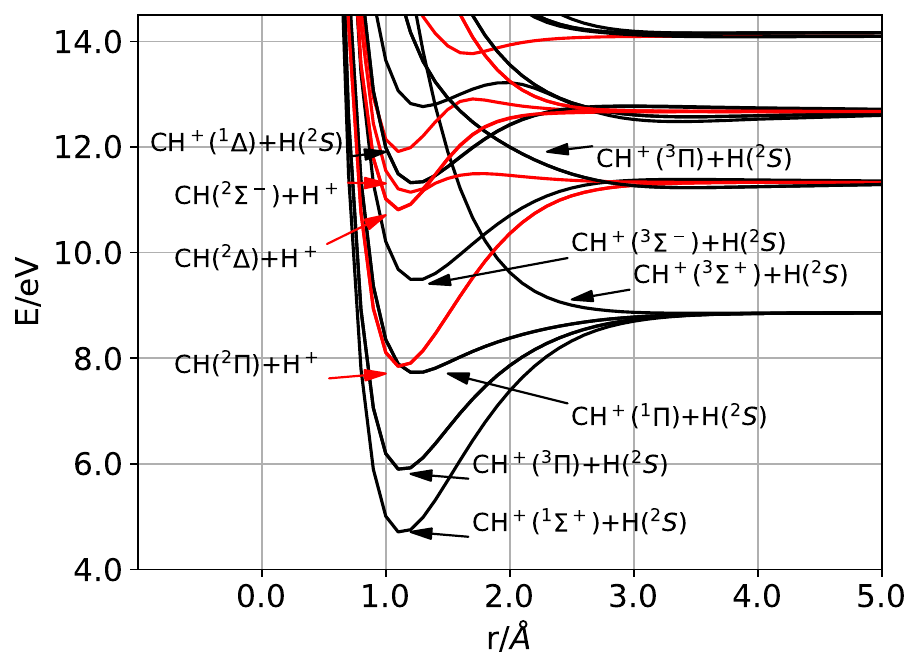}
    \includegraphics[width=.8\linewidth]{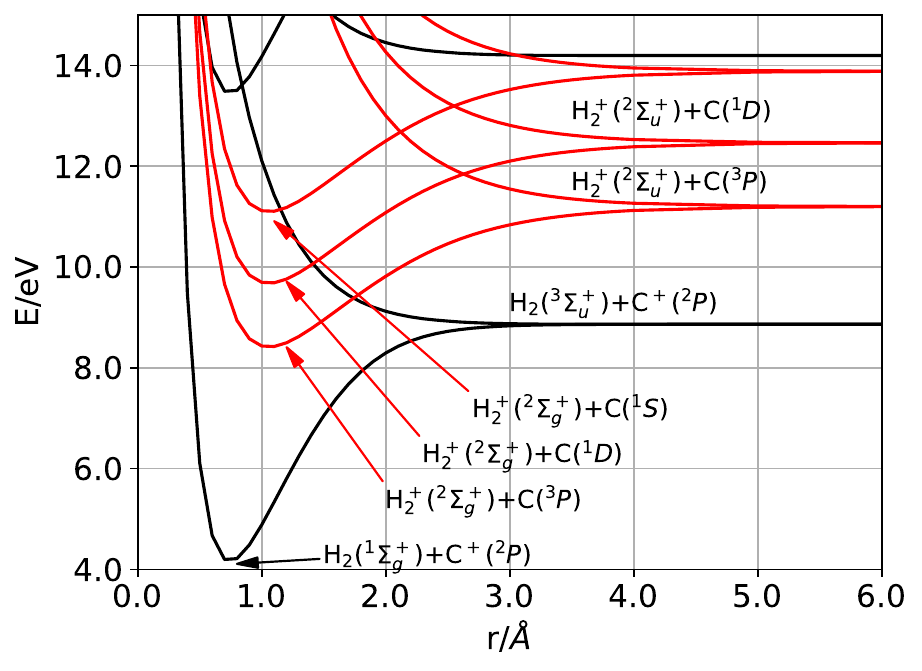}
    \caption{Top panel: Diatomic diabatic potential energy curves for the \ce{[CH + H]+} asymptote.
      Bottom panel: Diatomic diabatic potential energy curves for the \ce{[C + H2]+} asymptote.
      The zero of energy is placed at the minimum of the \ce{CH2+} in the ground electronic state $X^2A'$.
    \label{fig:diatomics_abinitio}}
\end{figure}

The diatomic fragments in the \ce{[CH + H]+} and \ce{[C + H2]+} channels in Fig.~\ref{fig:diatomics_abinitio} present a great variety of electronic states, bound and dissociative, in which the diatomic fragment can be neutral or ionized. When performing the calculations for the whole triatomic system in the $A'$ or $A''$ representations the adiabatic states present many crossings, related for instance with charge transfers between the distant partners. These adiabatic states are first diabatized by performing independent calculations of each diatomic fragment, with different charges and in different irreducible representations. Fig.\ref{fig:diatomics} displays the diabatized diatomic curves for both \ce{CH/CH+} and \ce{H2 /H2+} systems.

\subsection{Correlation diagram}

The different electronic states in all the rearrangement channels are connected with the lowest energy total fragmentation asymptotes: \ce{C+}$(^2P)$ + H$(^2S)$ + H$(^2S)$, \ce{C}$(^3P)$ + H$(^2S)$ + H$^+$ and \ce{C}$(^1D)$ + H$(^2S)$ + \ce{H+}. A full description of the 22 states that correlate to these asymptotes would require a matrix size which is intractable and computationally very demanding to perform exact quantum dynamics simulations. Instead of including the full set of 10 electronic states correlating to the \ce{C}$(^1D)$ asymptote, we have only considered the 4 states corresponding to the two $\Delta$ components. This truncation will yield an incomplete description of some asymptotic states as we will discuss later. 

\begin{figure}[H]
    \centering
    \includegraphics[width=\linewidth]{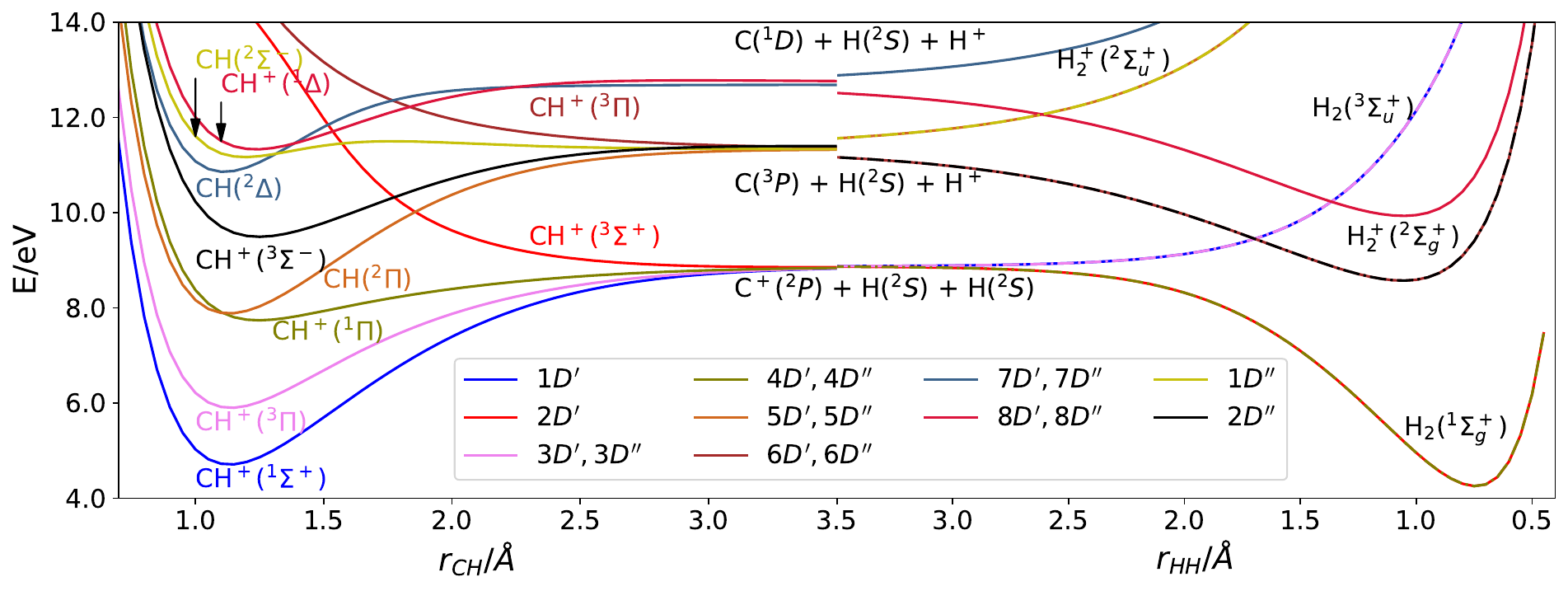}
    \caption{Representation of the diabatic diatomic curves included in the $\boldsymbol{U}_0$ model and correlation with the total dissociation asymptotes.
    \label{fig:diatomics}}
\end{figure}

Considering that the $\Lambda$ quantum number and charge of each diatomic fragment are conserved across the diabatic states, a correlation between reactants and products can be done. On top of that, it is necessary to use three--body information to complete the correlation of states with the same $\Lambda$ that tend to the same asymptote. The correlation diagram of the diatomic states is presented in Fig.~\ref{fig:diatomics}.

The diabatic states are ordered as follows. The first two states are $\Sigma$ and are different in the $A'$ and $A''$ diabatic manifolds. The next four states are of $\Pi$ character and are degenerate in the $A'$ and $A''$ manifolds, correlating to the same products. The same occurs for the last two states, which are of $\Delta$ character. In the following we will use $A'$ and $A''$ to refer to adiabatic states and $D'$ and $D''$ for diabatic states. The states included in the model are enumerated in Tab.~\ref{tab:relative_energies}, together with the energies in the different minima and total dissociation.
\begin{table*}[hbpt]
    \centering
    \begin{tabular}{ccrcrcr}
    \toprule
    State & \ce{[CH + H]+} & $\Delta E$/eV & \ce{[C + H2]+} & $\Delta E$/eV & \ce{[H + H + C]+} & $\Delta E$/eV \\ 
    \midrule
     $1 D'$ & \ce{CH+}($^1\Sigma^+$) + H($^2S$) & 4.71 & \ce{H2}($^3\Sigma_u^+$) + \ce{C+}($^2P$) & $-$ & H + H + \ce{C+}($^2P$) & 8.87\\
     $2 D'$ & \ce{CH+}($^3\Sigma^+$)  +  H($^2S$) & $-$	& \ce{H2}($^1\Sigma_g^+$) +  \ce{C+}($^2P$) & 4.25 &	H + H + \ce{C+}($^2P$) &8.87 \\
  $1D''$ & CH($^2\Sigma^-$) + \ce{H+} & 11.17 & \ce{H2+}($^2\Sigma_u^+$) + C($^3P$) & $-$ & \ce{H+} + H + C($^3P$) & 11.33 \\
  $2D''$ & \ce{CH+}($^3\Sigma^-$) + H($^2S$) & 9.49 & \ce{H2+}($^2\Sigma_g^+$) + C($^3P$) & 8.57 & \ce{H+} + H + C($^3P$) & 11.33\\
     $3D', 3D''$ & \ce{CH+}($^3\Pi$) + H($^2S$) & 5.90 & \ce{H2}($^3\Sigma_u^+$) + \ce{C+}($^2P$) & $-$ & H + H + \ce{C+}($^2P$) & 8.87\\
    $4D', 4D''$ & \ce{CH+}($^1\Pi$) +  H($^2S$) & 7.74 & \ce{H2}($^1\Sigma_g^+$) + \ce{C+}($^2P$) & 4.25 & H + H + \ce{C+}($^2P$) &	8.87 \\
    $5D', 5D''$ & \ce{CH}($^2\Pi$) + \ce{H+} & 7.88 & \ce{H2+}($^2\Sigma_u^+$) + C($^3P$) & $-$ & \ce{H+} + H + C($^3P$) & 11.33 \\
    $6D', 6D''$ & \ce{CH+}($^3\Pi$) + H($^2S$) & $-$ & \ce{H2+}($^2\Sigma_g^+$) + C($^3P$) & 8.57 & \ce{H+} + H + C($^3P$) & 11.33 \\
   $7D', 7D''$ & \ce{CH}($^2\Delta$) + \ce{H+} & 10.86 & \ce{H2+}($^2\Sigma_u^+$) + C($^1D$) & $-$ & \ce{H+} + H + C($^1D$) & 12.68 \\
   $8D', 8D''$	& \ce{CH+}($^1\Delta$) +  \ce{H}($^2S$) & 11.33 & \ce{H2+}($^2\Sigma_g^+$) + C($^1D$) & 9.94 & \ce{H+} + H + C($^1D$) & 12.68 \\
    \bottomrule
    \end{tabular}
    \caption{Minimum energies of the different diatomics and total dissociation channels included in the PEM.
      The energies are referred to the \ce{CH2+} minimum. Each line represents a diabatic state in the PEM.
      When no number is indicated the electronic state does not have a minimum.
    \label{tab:relative_energies} }
\end{table*}
\subsection{Potential energy matrix}
Two independent PEMs are considered: for the $A'$ and $A''$ irreducible representations. The diagonal elements correspond to the diabatic states and the non--diagonal elements to their couplings. The $\Pi$ and $\Delta$ subblocks are identical in both representations, as well as $\Pi-\Delta$ couplings, so the differences between the energies are only due to the $\Sigma$ states and how they couple with the $\Pi$ and $\Delta$ states in each representation.

In our model we describe 16 electronic states, 8 in each $A'$ and $A''$ representations. The PEM is factored into a zeroth order diagonal matrix ($\boldsymbol{U}^0$), which accurately includes all the possible diatomic and total dissociation states, plus a three--body neural network matrix to incorporate corrections to the diabatic states and their couplings in the 3--body region ($\boldsymbol{U}^{NN}$),

\begin{equation}
    \boldsymbol{U} = \boldsymbol{U}^0 + \boldsymbol{U}^{NN}
    \label{eq:PEM}
\end{equation}

The elements of $\boldsymbol{U}^0$ are expressed as:
\begin{equation}
    U^0_{ii} = V_{1B} + V_{2B} + V_{lr}
\end{equation}
where $V_{1B}$ are the atomic energies of the total dissociation asymptote, $V_{2B}$ is the sum of the three possible diatomics in the diabatic state and correspond to diabatized states of the \ce{CH+}, CH, \ce{H2+} or \ce{H2} systems. $V_{lr}$ is the long range interaction of the rearrangement channel.

The $\boldsymbol{U}^{NN}$ matrix elements are computed from a permutationally invariant polynomial neural network (PIP--NN)~\cite{Jiang-Guo:13}. Those associated to $\Pi$ and $\Delta$ states are shared between the two PEM in order to maintain the degeneracy in linear configurations. The output of the neural network is multiplied by a damping function of the CH distances to make them zero as the system tends to a 2--body or total dissociation region. Additional constraints to the non--diagonal elements arise from the fact that for highly symmetric configurations ($C_{\infty v}, D_{\infty h}$) the PEM should resemble the adequate block structure. To impose these constraints, couplings between $\Sigma-\Pi$, $\Sigma-\Delta$ and $\Pi-\Delta$ are multiplied by a sine function of the H--C--H angle. The full diabatic matrices and further details on the neural network implementation are presented in the SI. 

{
	While in theory this method can be generalized to more than 3 particles, the main difficulty resides in the definition of $\boldsymbol{U}^0$. As the number of atoms increases, the number of relevant asymptotic states can be very large, which would lead to an intractable diabatic matrix. On the other hand, $\boldsymbol{U}^{NN}$ can be easily generalized to other systems with a PIP representation~\cite{JLi2020,delMazo-Sevillano-etal:24a,Houston2024}.
}

$\boldsymbol{U}^{NN}$ is trained by minimizing the following loss function ($\mathcal{L}$) which, in contrast to other methods, only requires adiabatic information and hence makes the { diabatization} method automatic:
\begin{equation}
    \mathcal{L} = \frac{1}{n_a N}\sum_{s}^{n_a}(\mathbf{V}_s - \mathbf{V}_{s,pred})^2 + \frac{\lambda_2}{n_d N}\sum_{s}^{n_d}\sum_{t\ne s}^{n_d}(\mathbf{U}_{st,pred})^2 \label{eq:loss}
\end{equation}
where $\mathbf{V}_s$ and $\mathbf{V}_{s,pred}$ are the {\it ab initio} and predicted adiabatic energies for the electronic state $s$. The second term regularizes the training process forcing the non--diagonal terms to be small and avoiding too large couplings where not needed. In this case $\lambda_2 = 10^{-3}$
{
so that non--diagonal couplings are only minimized if they do not affect the accuracy of the adiabatic energies}. $N$ is the total number of {\it ab initio} energies and $n_a$ and $n_d$ the number of adiabatic and diabatic states which are trained in each representation, respectively.

Two key aspects make this method different from similar PEM factorizations such as the PM-DDNN~\cite{shuParametricallyManagedActivation2024} and turn the { diabatization} method into fully automatic: 1) sharing the degenerate diabatic states and couplings among the $A'$ and $A''$ irreducible representations and 2) constraining the non--diagonal couplings at highly symmetric configurations to resemble the adequate $\Sigma$, $\Pi$ and $\Delta$ subblock structure in linear configurations.

\subsection{Features of the PEM}
The PEM has been trained with $n_a=5$ and $n_d=8$ in Eq.~\eqref{eq:loss}, meaning that the neural network term will modify the 8 diabatic states in each representation to finally reproduce accurately the 5 lowest adiabatic states of each representation. These states cover an energy up to 14 eV in the Frank-Condon region of the \ce{CH2+}, which is the relevant energy range for astrochemical considerations. In the diatomic and total dissociation regions the number of accurate states is in general larger than five thanks to the $\boldsymbol{U}_0$ term. The {\it ab initio} calculations have been performed with the state averaged CASSCF level of theory followed by a MRCI calculation with the cc-pCVTZ basis set with ORCA 6.0~\cite{RN269, RN176, RN232, RN243}. In all the calculations the core orbitals of the C atom are closed. A total of seven doublet states are computed for each representation of the $C_{s}$ point group.

\begin{figure}[H]
    \centering
    \includegraphics[width=\linewidth]{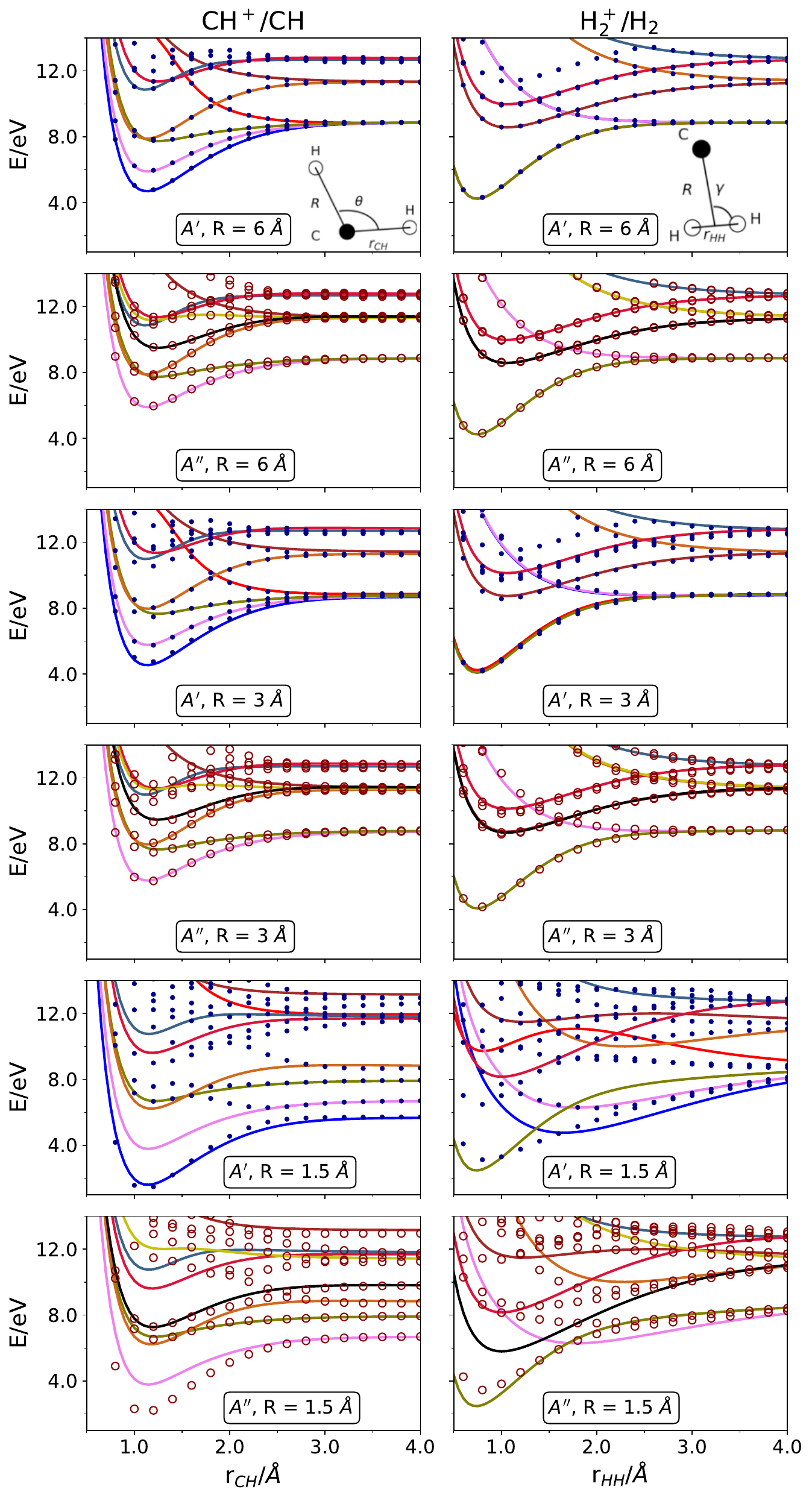}
    \caption{Black dots represent the {\it ab initio} energy and solid lines the diabatic energies predicted by $\boldsymbol{U}^0$.
      From top to bottom, represents the change of the diatomic curves as the third atoms approaches. In all cases $\theta=140^\circ$ and $\gamma=90^\circ$.
    \label{fig:U0_approach} }
\end{figure}

$\boldsymbol{U}^0$ is responsible of reproducing the full PEM up to a rather small distance between the three atoms as depicted in Fig.~\ref{fig:U0_approach}, where a comparison between $\boldsymbol{U}^0$ and the {\it ab initio} data is presented. For $R=6$ \AA\, we find a good agreement with the exception of the states that are not included in the zeroth order model: some that correlate with C($^1D$) or the curve corresponding to \ce{H2} + \ce{C+}($^4P$) which is visible in the right panels. As the third atom approaches to distances $\approx 3$ \AA\,, $\boldsymbol{U}^0$ is still rather accurate and only for very short distances the zeroth order model is clearly in disagreement. It is in this short range region where $\boldsymbol{U}^{NN}$ acts, correcting the $\boldsymbol{U}^0$ diabatic energies and including the missing couplings.

\begin{figure}[H]
    \centering
    \includegraphics[width=\linewidth]{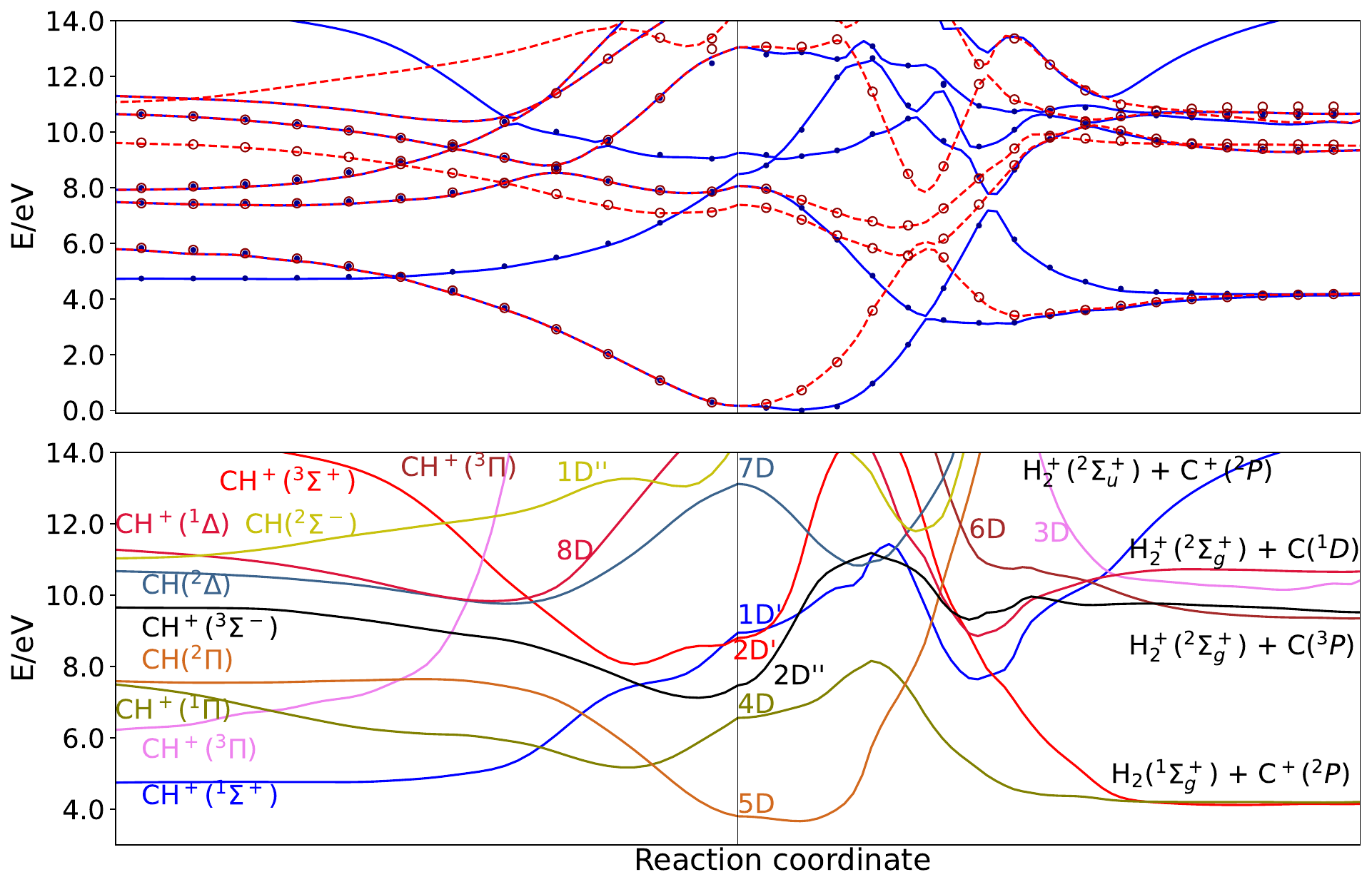}
    \caption{Representation of a path joining the \ce{[CH + H]+} and \ce{[C + H2]+} asymptotes from the \ce{CH2+} linear configuration in the center of the plot.
      Towards the \ce{[CH + H]+} asymptote a C--H distances is elongated. Towards the \ce{[C + H2]+} asymptote the extraction of the C between both H
      atoms is coordinated with a progressive decrease of the H--H distance towards its equilibrium geometry in the \ce{H2} molecule.
      In the top and bottom panels the adiabatic and diabatic energies are presented, respectively. The dots in the top panel represent the {\it ab initio}
      energies for those geometries.
    \label{fig:mep} }
\end{figure}

Fig.~\ref{fig:mep} displays a path connecting the \ce{CH2+} in its linear configuration with the \ce{[CH + H]+} asymptote on the left and the \ce{[C + H2]+} asymptote in the right. In the top panel the adiabatic energies are compared with the calculated {\it ab initio}, represented with dots. For the 5 lowest adiabatic states we find a very good agreement along the path. In the bottom panel, the evolution of the diabatic states towards the photofragments is depicted. For instance, the minimum in the \ce{CH2+} configuration comes from the $5D', 5D''$ diabatic states which connect \ce{CH}($^2\Pi$) + \ce{H+} and \ce{H2+}($^2\Sigma_u^+$) + C($^3P$) asymptotes, in accordance with our previous correlation diagram for this system~\cite{delMazo25_etal_CHp}. However, this is not evident from Fig.~\ref{fig:U0_approach}. In $\boldsymbol{U}^0$ the lowest diabatic $\Pi$ state close to the \ce{CH2+} minimum are the $3D', 3D''$ states, which correlate with \ce{CH+}($^3\Pi$) + H($^2S$) and \ce{H2}($^3\Sigma_u^+$) + \ce{C+}($^2P$), so one could assume that $\boldsymbol{U}^{NN}$ will try to preserve this order. Instead, $\boldsymbol{U}^{NN}$ pushes the $3D',3D''$ states up in energy while bringing down the $5D', 5D''$ to reproduce the minimum of the \ce{CH2+}. 
Further comparisons of the adiabatic and {\it ab initio} results can be found in the SI. 

\begin{table}[H]
    \centering
    \begin{tabularx}{\linewidth}{cXXXXX}
    \toprule
        Irrep. & 1 & 2 & 3& 4 & 5\\
     \midrule
         $A'$&  35.0 (20542) & 41.3 (20209) & 46.1 (19557) & 76.4 (18763) & 523.8 (18331)\\
         $A''$& 45.4 (20421) & 43.7 (19963) & 41.2 (19239) & 57.0 (18500)  & 519.2 (17682)\\
         \bottomrule
    \end{tabularx}
    \caption{RMSE for the five lowest adiabatic states in each representation. The errors are given in meV and in parenthesis
      the number of points with energy up to 14 eV with respect to the \ce{CH2+} minimum, which have been used to compute the error.
    \label{tab:rmse} }
\end{table}

The root mean squared errors (RMSE) of the different states are presented in Tab.~\ref{tab:rmse}. The largest error in the fifth state is due to the missing $\Sigma$ and $\Pi$ components of the \ce{C}$(^1D)$ + H$(^2S)$ + \ce{H+} asymptote, which limits the capability of the model to describe the \ce{H2+}($^2\Sigma_g^+$) + C($^1D$) asymptote . Excluding \ce{[C + H2]+} geometries, the errors of the fifth adiabatic state up to 14 eV are 86.0 meV and 76.4 meV for the $A'$ and $A''$ irreducible representations, respectively.

\begin{figure}[H]
    \centering
    \includegraphics[width=.8\linewidth]{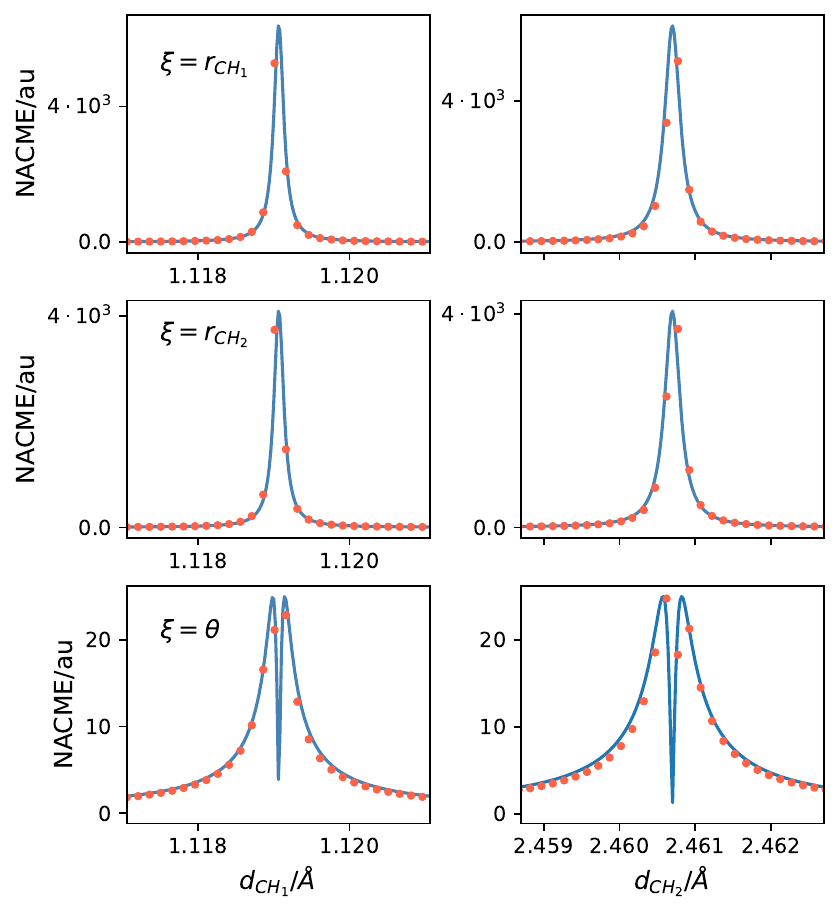}
    \caption{Non--adiabatic coupling matrix elements computed close to the CI ($\theta=0.01^\circ$) between the first two adiabatic states in
      the \ce{CH+ + H} exit. In blue those from the PEM and in red the {\it ab initio} computed at a CASSCF level.
      $\xi$ represents the coordinate with respect to which the NACME is computed.
    \label{fig:nacme} }
\end{figure}

More relevant to the quantum dynamics are the non--adiabatic coupling matrix elements (NACMEs) which can be analytically computed from the PEM and compared with {\it ab initio} calculations in regions close to a CI~\cite{Sanz-Sanz-etal:15}. Fig.~\ref{fig:nacme} presents a comparison between those computed from the PEM and {\it ab initio} at a CASSCF level close to the CI between the first two adiabatic states in the \ce{CH+ + H} exit.

\section{Photodissociation dynamics}

\begin{figure}[H]
    \centering
    \includegraphics[width=.95\linewidth]{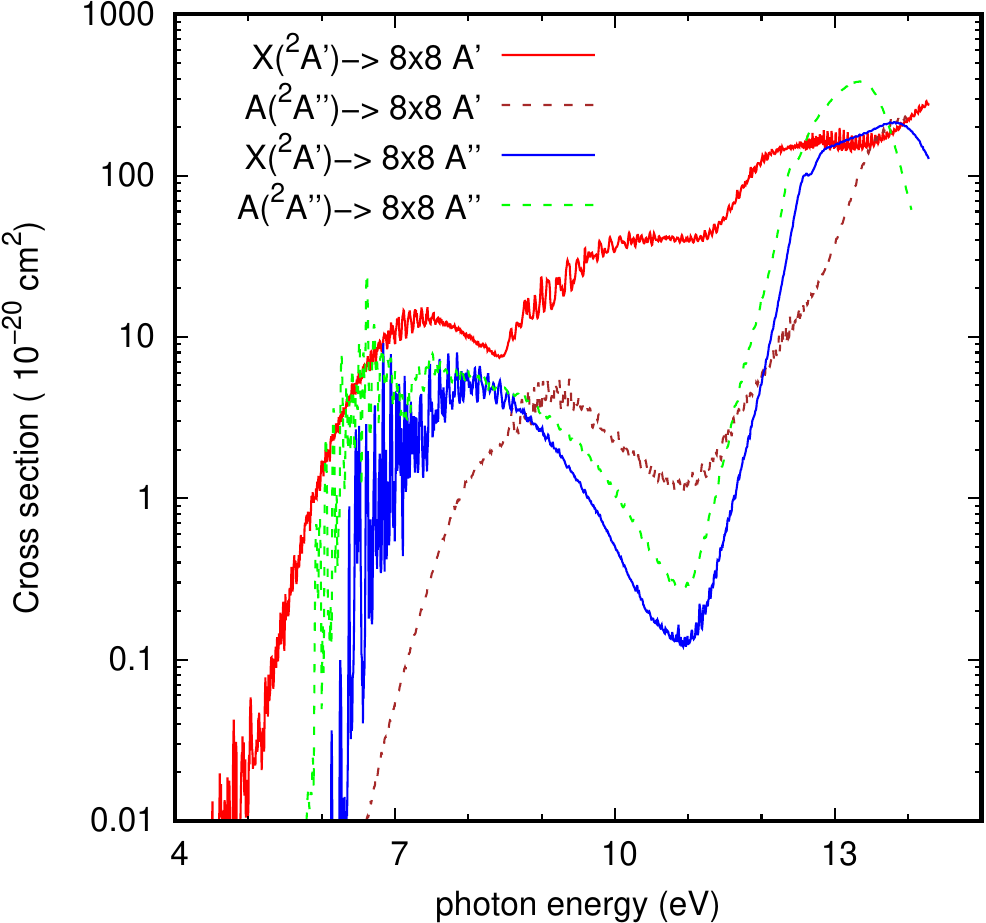}
    \caption{Photodissociation cross sections for the  $J_i=0\longrightarrow J=1^-$  transition from the $\widetilde{X} ^2A'(0,0,0)$ (solid lines) panels)
      or $\widetilde{A} ^2A''$(0,0,0) (right panels) to the 8$\times$8 $A'$ (bottom panels) or $A''$ (dashed lines) coupled diabatic states.
    \label{fig:8x8photodissociationXsection}}
\end{figure}

The photodissociation dynamics is studied with the quantum time-dependent MADWAVE3 code\cite{Roncero-delMazo-Sevillano:25}
as described previously  \cite{Paniagua-etal:99,Aguado-etal:03,Chenel-etal:16,Aguado-etal:17} using the CH+H  Jacobi coordinates, with {\bf r}
being the HC internuclear vector, {\bf R} joining HC center-of-mass to the second H atom, and $\gamma$=${\bf r}\cdot{\bf R}/rR$. The rovibrational H$_2$ products are analyzed using the reactant-product transformation method \cite{Gomez-Carrasco-Roncero:06}.
The dynamics is performed in the 8$\times$8 diabatic representations of the $A'$ and $A''$ symmetries independently.

Here we consider the absorption from the ground vibrational states in the adiabatic ${\widetilde X}^2 A'$ and ${\widetilde A}^2A''$ states, described in the SI, corresponding to initial total angular momentum $J_i$ =0. The initial wave packet is built by multiplying the bound state by the transition electric dipole moment as described previously \cite{Paniagua-etal:99,Aguado-etal:03,Chenel-etal:16,Aguado-etal:17}. The transition dipole moments
have been calculated from the ground adiabatic states to the 5 (4) adiabatic excited electronic $A'$ ($A''$), as described in the SI.
These adiabatic transition dipole moments are then transformed to the diabatic representation using the eigen-vectors obtained after diagonalization of the $8\times 8$ diabatic matrix at each nuclear configuration. Thus, the simultaneous excitation of the first 5 (4) $A'$ ($A''$) adiabatic states  is considered here.

We study the $J_i=0$ $\rightarrow$ $J=1^-$ rotational excitation for 4 different electronic transitions: from either $\widetilde X^2A'(0,0,0)$ or $\widetilde A ^2A''(0,0,0)$ initial states towards  8 electronically excited and coupled diabatic states, of either $A'$ or $A''$ symmetry. 

The absorption spectra obtained from the autocorrelation function\cite{Chenel-etal:16,Aguado-etal:17} shows an excellent agreement
with the total photodissociation flux over all fragments above the first dissociation limit, as displayed in Fig.~\ref{fig:8x8photodissociationXsection}. This serves not only to check to convergence of the calculation, but also to guarantee that all the flux over different fragmentation channels is properly collected. 

The assignment of the bands is done by comparing these spectra to those obtained in the adiabatic
representation, as described in the SI. The most intense peaks are those arriving to 5$A'$ and 4$A''$ electronic states, because these excited states have considerably larger  transition dipole moments with $\widetilde X ^2A'$ and $\widetilde A ^2A''$ states than the other electronic states at lower energies.

The  spectra to  the $A'$ manifold of states obtained from the ${\widetilde X}$ or the ${\widetilde A}$  states show significant differences,  attributed to the different initial wave packets, formed from different initial bound states (specially in the angular coordinate) and different transition dipole moments (see SI). 

The transitions towards the $A''$ states show a drop around 10 eV, which is clearly attributed to the energy separation among the $A''$ states, less densely packed than the $A'$ states. The drop at 10 eV observed in the $\widetilde A ^2A'' \longrightarrow$ $A'$ transition  is due to the reduction of the 1$A''$$\longrightarrow$ 3 and 4 $A'$ matrix elements of the dipole moments.

\begin{figure*}[t]
    \centering
    \includegraphics[width=.45\linewidth]{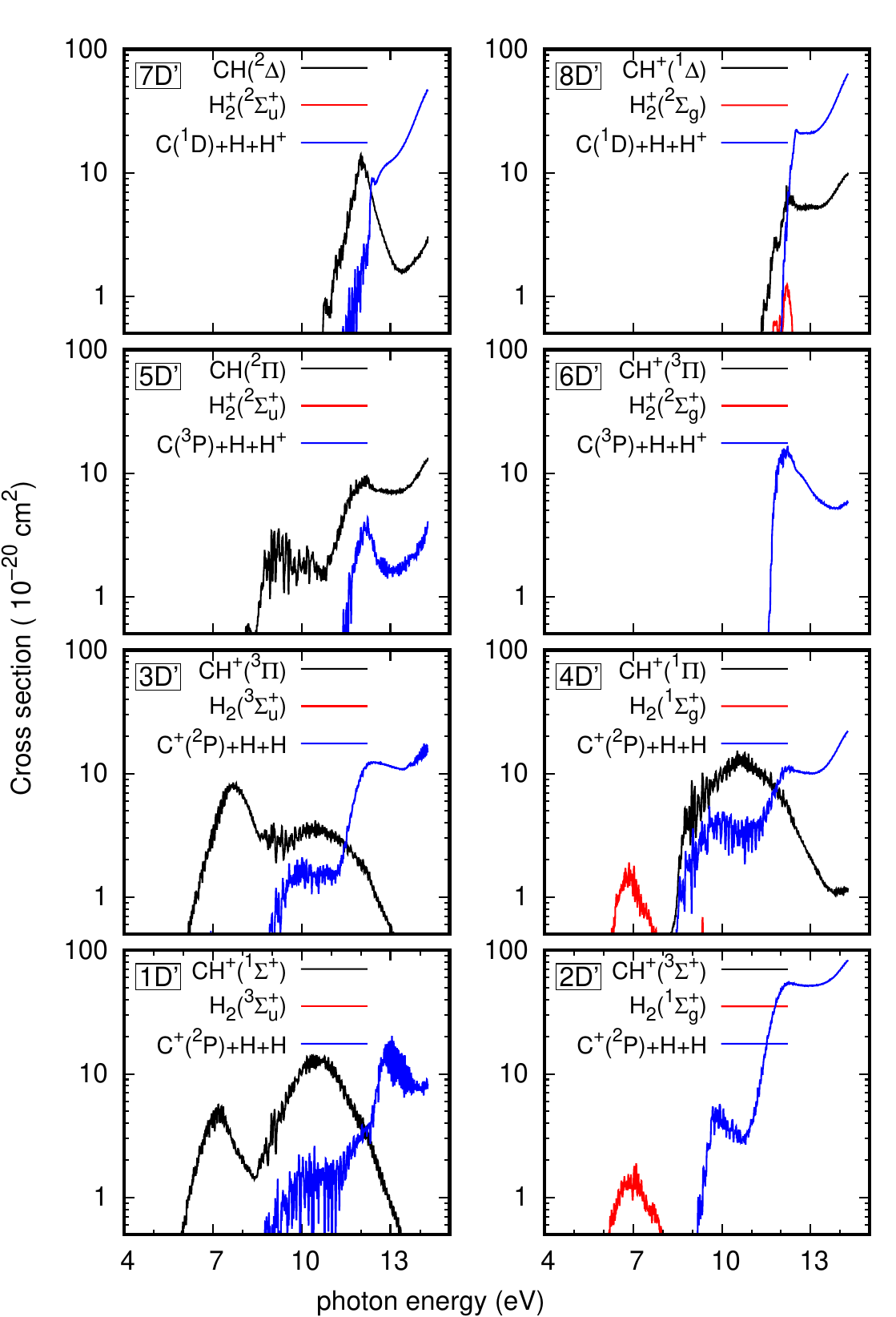}
    \includegraphics[width=.45\linewidth]{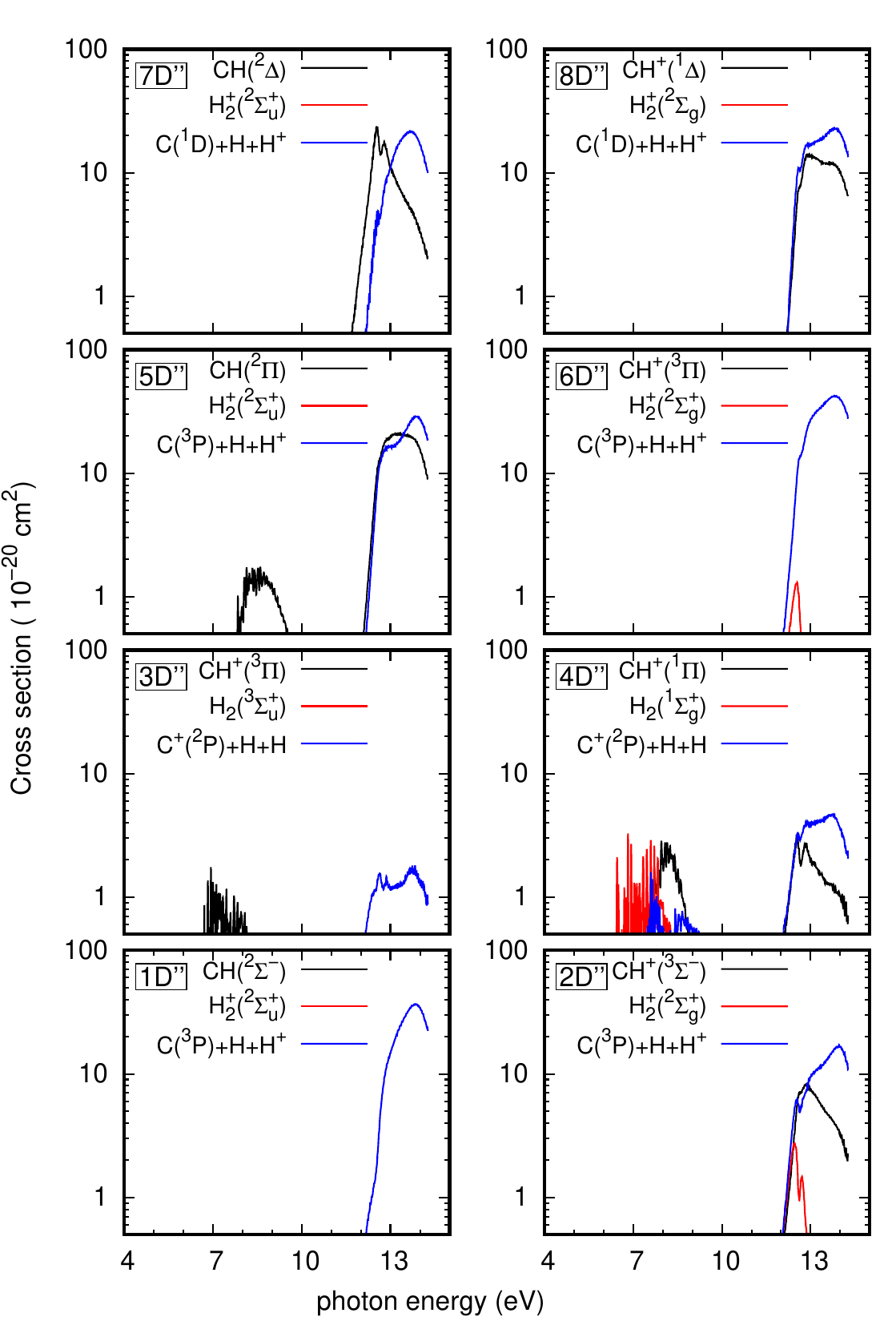}
    
    \caption{Photodissociation cross sections for the $\widetilde X ^2A'$(0,0,0) $\longrightarrow$ all $D'$ states  and all $D''$ states, as indicated in each panel, separating individual fluxes for the different fragments of each rearrangement channel in each electronic state, as discussed in the text. 
    \label{fig:electronicFluxes-8x8ApYAsfromAp} }
\end{figure*}

There is a great variety of fragmentation processes, as illustrated by the electronic states of the fragments shown in Fig.~\ref{fig:diatomics}: total fragmentation
\begin{eqnarray}\label{eq:total-fragmentation}
    \ce{CH2+}(\widetilde X^2A', \widetilde A^2A'') +h\nu &\longrightarrow& \ce{C^+ + H + H}\\ 
                          &\longrightarrow& \ce{C + H^+ + H}\nonumber,
\end{eqnarray}
 formation of CH or CH$^+$ 
\begin{eqnarray}\label{eq:chYchp-formation}
    \ce{CH_2^+}(\widetilde X^2A', \widetilde A^2A'') +h\nu &\longrightarrow& \ce{CH^+  + H}\\ 
                          &\longrightarrow& \ce{CH + H^+} \nonumber,
\end{eqnarray}
and formation of H$_2$ or H$_2^+$
\begin{eqnarray}\label{eq:h2Yh2p-formation}
    \ce{CH_2^+}(\widetilde X^2A', \widetilde A^2A'') +h\nu &\longrightarrow& \ce{H_2  + C^+}\\ 
                          &\longrightarrow& \ce{H_2^+ + C}\nonumber.
\end{eqnarray}
These different fragmentation routes are distinguished by the analysis of different fluxes on each electronic state and  on individual rovibrational states of H$_2$/H$_2^+$ and CH/CH$^+$ diatomic products\cite{Gomez-Carrasco-Roncero:06}. The total flux, obtained for the CH internuclear distance $r$= 6 \AA, collects the total fragmentation, Eq.~\ref{eq:total-fragmentation}, the formation of H$_2$ or H$_2^+$, Eq.~\ref{eq:h2Yh2p-formation},  depending on the electronic state, and one half of CH or CH$^+$, Eq.~\ref{eq:chYchp-formation}, expressed as
\begin{eqnarray}
  F_{total}= F_{\ce{H + H + C}}+F_{\ce{H_2 + C}} + {1\over 2} F_{\ce{CH + H}} .
\end{eqnarray} 

{
The fluxes for  H$_2$/H$_2^+$ and one of the two CH/CH$^+$ rearrangement channels are obtained by evaluating the sum over all possible rovibrational diatomic eigen states in each channel for each electronic states, giving the quantities $F_{\ce{H_2 + C}}$ and $F_{\ce{CH + H}}/2$. For the \ce{[CH + H]+} there are two equivalent rearrangement channels giving the same flux,  $F_{\ce{CH + H}}/2$ each.
}
 In Figs.~\ref{fig:electronicFluxes-8x8ApYAsfromAp} the cross sections for each channel and final electronic states obtained after photodissociation of the $\widetilde X ^2A'(0,0,0)$ ground state towards the 8 lower $A'$ and $A''$ final  electronic states, respectively, are shown with a similar qualitative behavior to those obtained from $\widetilde A ^2A''(0,0,0)$ initial state shown in the SI.

The main fluxes in the higher energy interval, $E>11$ eV, are either to total fragmentation or to CH/CH$^+$ diatomic fragments, with small proportion of H$_2$/H$_2^+$ products. For the 5,6,7 and 8 diabatic $D'$ or $D''$ states, the total fragmentation yield neutral carbon atoms, in either $^3P$ or $^1D$ states and the charge is in the hydrogen atom. On the contrary, for the 3,4 $D'$ or $D''$ states the charge is in the carbon atoms in the C$^+$($^2P$) state. The $\Sigma$ states in the $A'$ and $A''$ symmetry show different behavior: in the $A'$ manifold the charge is in the C$^+(^2P)$ ion, while in the $A''$ manifold the charge is in the hydrogen atoms, leading neutral C($^3P$) atoms.

The most astonishing result is the high cross section to form neutral CH products, in either the ground $X^2\Pi$ (states 5 $D'$ and 5 $D''$) or $A^2\Delta$ (states 7$D'$ and 7$D''$). CH can be formed in the neutral-neutral reaction C($^3P)$+H$_2$, but is endothermic by $\approx$ 1 eV. Thus the neutral reaction can only occur when H$_2$ is initially vibrationally excited in $v$ = 2 or 3. Abundant highly vibrationally excited H$_2$ is only found in strongly UV illuminated regions such as  the borders of molecular clouds (photodissociation or PDR regions) or protoplanetary discs (PPDs). In such environments, the high UV flux will also induce the formation of neutral CH fragments by the photodissociation of CH$_2^+$.

The formation of CH$_n^+$  is attributed to the 
 hydrogenation chain of reactions  CH$_n^+$+H$_2$ $\longrightarrow$ CH$_{n+1}^+$ + H, which stops in CH$_3^+$, which in turn reacts with many other molecules, being the origin of several chemical networks for the formation of many larger hydrocarbons. 
 CH$^+$ is one of the first observed molecules in the ISM \cite{Douglas-Herzberg:41} and since then widely observed. CH$_3^+$ was recently observed \cite{Berne-etal:23} in a protoplanetary disk (d203-506) illuminated by the strong far ultraviolet (FUV) radiation field from nearby massive stars in Orion's trapezium cluster. However, CH$_2^+$ has not been observed so far.

 Thus the confirmation of this hydrogenation chain requires the detection of CH$_2^+$, and its abundance determination requires incorporating all the destruction pathways, such as photodissociation, dissociative recombination with electrons and hydrogenation.

\begin{figure}[H]
    \centering
    \includegraphics[width=.7\linewidth]{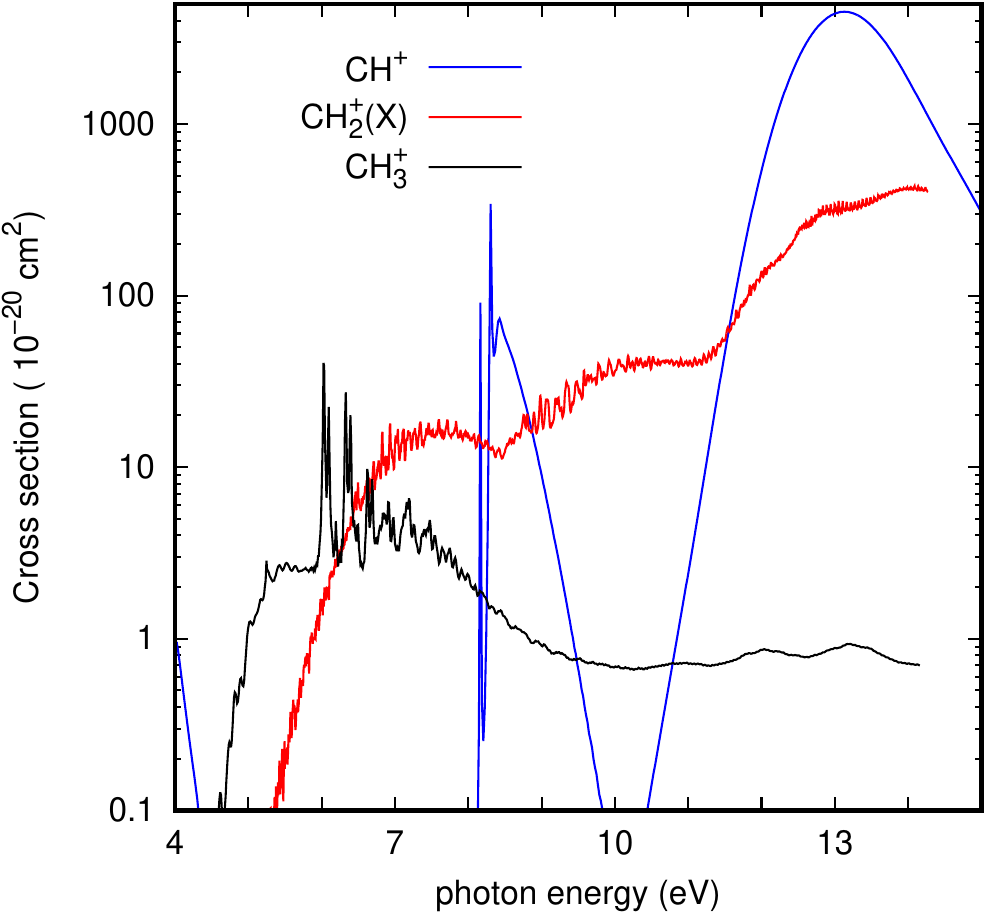}
    \caption{Total photodissociation cross section for the ground vibrational state of CH$_2^+$ ($\widetilde X ^2A')$
      (summed over all $A'$ and $A''$ states), compared to those of CH$_3^+$ from Ref.\cite{delMazo-Sevillano-etal:24},
      and  CH$^+$, recalculated here (see text) and very similar to that of Kirby  {\it et al.}\cite{Kirby-etal:80}.
    \label{fig:total-photodissociationXsections}}
\end{figure}

The total photodissociation cross section of CH$_2^+(\widetilde X)$ (summing the contributions towards $A'$ and $A''$ manifolds) is shown in Fig.~\ref{fig:total-photodissociationXsections} and compared to those of CH$_3^+$ from Ref.\cite{delMazo-Sevillano-etal:24} and CH$^+$, which is recalculated in this work as described in the SI and is in agreement with that reported by Kirby {\it et al. }\cite{Kirby-etal:80}. The photodissociation cross section of CH$_2^+$ is larger than that of CH$_3^+$ but smaller than that of CH$^+$. 

\begin{figure}[H]
    \centering
    \includegraphics[width=.9\linewidth]{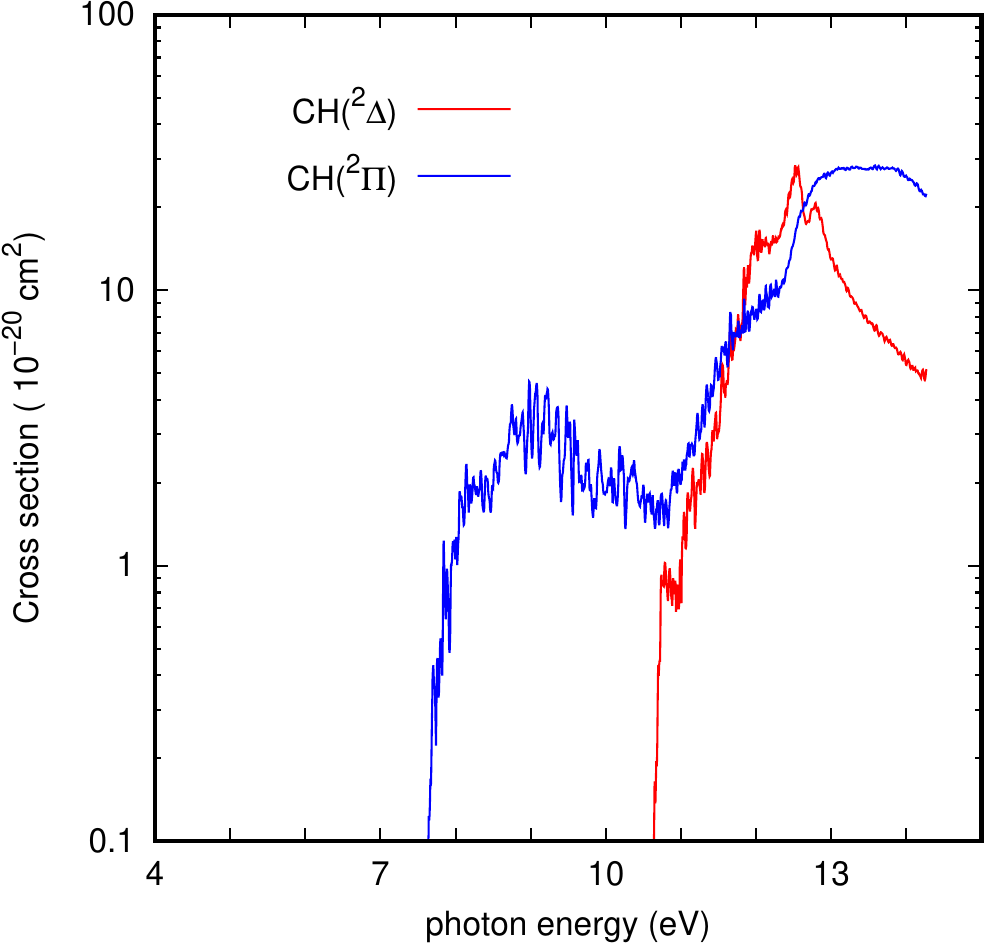}
    \caption{Total photodissociation cross section to form neutral CH products from the ground vibrational state of CH$_2^+$ ($\widetilde X^2A')$  (summed over all $A'$ and $A''$ states)
    \label{fig:CHtotal-photodissociationXsections}}
\end{figure}

The photodestruction is determined by the photodissociation rate obtained as the integral of the photodissociation cross section with the energy dependent radiation field. Using Draine's mean interstellar radiation field \cite{Draine:78},
the photodissociation rate of CH$_2^+$ is 6.45 $\times 10^{-11}$ s$^{-1}$  , while that of CH$_3^+$ is\cite{delMazo-Sevillano-etal:24}  6.83 $\times 10^{-12}$ s$^{-1}$ and for CH$^+$ is 
2.89 $\times 10^{-10}$ s$^{-1}$. The rates obtained here for CH$_2^+$ and CH$^+$ are slightly lower than the values reported for the ISRF field in Ref.\cite{Heays-etal:17} of 1.4  and 3.3 $\times 10^{-10}$ s$^{-1}$,  respectively. One reason is that  the ISRF flux is a modification of  the Draine's field. The difference for CH$_2^+$ is more important because a simple vertical transition was assumed to calculate the photodissociation cross section.
The total photodissociation cross section to form neutral CH products is shown
in Fig.~\ref{fig:CHtotal-photodissociationXsections}, which are rather high specially
for CH(X$^2\Pi$) at higher photon energies, where the UV radiation field is not shielded. The photodissociation rates to form CH in the $^2\Pi$ and $^2\Delta$ electronic states are 9.96 $\times 10^{-12}$ s$^{-1}$
and 3.37 $\times 10^{-12}$ s$^{-1}$, respectively, using Draine's mean interstellar radiation field \cite{Draine:78}. These values are small,
but considering that the H$_2$+C neutral reaction is endothermic by 1 eV,
it can be an alternative source of CH.

\section{Conclusions}
The study of the branching ratios among the different photoproducts in photodissociation over large energy intervals, as needed in astrochemistry, requires the description of non-adiabatic dynamics, involving many electronic states. To overcome the singularities of non-adiabatic couplings at conical intersections, the most widely used method is a unitary transformation to a diabatic representation, which is not unique. There are many diabatization methods, but usually they are restricted to few states. In this work we have developed a diabatization method describing up to 16 electronic states, divided in two symmetry blocks. The method has been applied to the photodissociation of CH$_2^+$, not yet dynamically studied, and that present many photoproducts, bound and dissociative and with several charge distributions.

The diabatization method consists in building a basis of electronic states to represent the possible products, with different charges and electronic excitation. A correlation among the rearrangement channels is done, imposing symmetry restrictions, such as the conservation of the projection of angular momentum, $\Lambda$, on the diatomics fragments. The full diabatic matrices are then factorized in two terms, a diagonal zero-order term and a full matrix built using a neural network method, $V_{NN}$. The zero-order term allows to describe the diagonal interaction among the fragments up to intermediate distances, $\approx$ 3-4 \AA, keeping the nature of the asymptotic states. The $V_{NN}$ matrix is build with an artificial neural network, imposing symmetry restrictions on the couplings at symmetric configurations and introducing a penalty function to force non-diagonal terms to be as small as possible. This process allowed an accurate description of the lowest 10 adiabatic eigenvalues in the short distance range and 16 in the asymptotes, separated in two symmetry blocks, covering energies up to 14 eV.

The diabatic model allows the quantum dynamics of the total photodissociation cross sections rate constants for CH$_2^+$ up to 13.6 eV, interval needed to astrochemical models. More important, it allows to describe the partial cross section for a great variety of fragmentation processes, with the products in different arrangement channels, charge and electronic states: total fragmentation, C$^+$+H+H and C+H+H$^+$, CH$^+$(X$^1\Sigma$, )+H, CH(X$^2\Pi$,$^2\Delta$), H$_2$(X$^1\Sigma$,)+C$^+$ and H$_2^+$($^2\Sigma_g$). 

Here we focus in the formation of the CH radical, whose formation in the C+H$_2$ reaction is endothermic. In strongly illuminated astronomical objects, the CH$_2^+$ photodissociation leads to significant of CH radical, and it is important to determine its contribution in PDRs and PPDs under different conditions to determine its contribution.

CH$_2^+$ is a prototype and this kind of studies are needed to search for alternative  formation routes of many molecules and radicals in astronomical objects illuminated by UV radiation. For this purpose diabatization is a key important requirement, specially in cases where the fragments present crossings.  At these crossings at long distance the non-adiabatic couplings vanish and surface hopping methods would adiabatically continue in one adiabatic state, corresponding to a mixture of two states each one assigned to one chemical product, as illustrated in Fig.~\ref{fig:diatomics} for the present case.

\section*{Author contributions}

\textbf{Pablo del Mazo-Sevillano}: Conceptualization (equal); Methodology (equal); Software (equal); Writing – review \& editing (equal). \textbf{Alfredo Aguado}: Conceptualization (equal); Funding acquisition (equal); Methodology (equal); Writing – review \& editing (equal). \textbf{Susana Gómez-Carrasco}: Conceptualization (equal); Funding acquisition (equal); Writing – review \& editing (equal). \textbf{Octavio Roncero}: Conceptualization (equal); Funding acquisition (equal); Methodology (equal); Project administration (equal); Software (equal); Writing – review \& editing (equal).

\section*{Conflicts of interest}
There are no conflicts to declare.

\section*{Data availability}

The total photodissociation cross sections calculated here for CH$_2^+$  and \ce{CH+} are available at ZENODO \url{https://zenodo.org/records/19847742}.
The codes used for the calculations have been properly cited, and
the details of the calculations are provided in the ESI. Any other data that support the findings of this study are available from the authors,  upon reasonable request. 

\section*{Acknowledgements}
This work has received funding from Ministerio de Ciencia,
Innovaci\'on y Universidades, MICIU (Spain), under Grants No.
PID2021-122549NB-C21, PID2021-122549NB-C22 and  PID2024-156686NB-I00. Computational
assistance was provided by the Supercomputer facilities of Lusitania founded by the C\'enitS and Computaex Foundation.



\balance



\begin{mcitethebibliography}{53}
\providecommand*{\natexlab}[1]{#1}
\providecommand*{\mciteSetBstSublistMode}[1]{}
\providecommand*{\mciteSetBstMaxWidthForm}[2]{}
\providecommand*{\mciteBstWouldAddEndPuncttrue}
  {\def\EndOfBibitem{\unskip.}}
\providecommand*{\mciteBstWouldAddEndPunctfalse}
  {\let\EndOfBibitem\relax}
\providecommand*{\mciteSetBstMidEndSepPunct}[3]{}
\providecommand*{\mciteSetBstSublistLabelBeginEnd}[3]{}
\providecommand*{\EndOfBibitem}{}
\mciteSetBstSublistMode{f}
\mciteSetBstMaxWidthForm{subitem}
{(\emph{\alph{mcitesubitemcount}})}
\mciteSetBstSublistLabelBeginEnd{\mcitemaxwidthsubitemform\space}
{\relax}{\relax}

\bibitem[van Hemert and van Dishoeck(2008)]{Hemert-vanDishoeck:08}
M.~C. van Hemert and E.~F. van Dishoeck, \emph{Chem. Phys.}, 2008,
  \textbf{343}, 292\relax
\mciteBstWouldAddEndPuncttrue
\mciteSetBstMidEndSepPunct{\mcitedefaultmidpunct}
{\mcitedefaultendpunct}{\mcitedefaultseppunct}\relax
\EndOfBibitem
\bibitem[van Dishoeck and Visser(2015)]{Dishoeck-Visser:15}
E.~F. van Dishoeck and R.~Visser, \emph{Laboratory Astrophysics: from Molecules
  through Nanoparticles to Grains}, Wiley-VCH, 2015, p.~1\relax
\mciteBstWouldAddEndPuncttrue
\mciteSetBstMidEndSepPunct{\mcitedefaultmidpunct}
{\mcitedefaultendpunct}{\mcitedefaultseppunct}\relax
\EndOfBibitem
\bibitem[Heays \emph{et~al.}(2017)Heays, Bosman, and van
  Dishoeck]{Heays-etal:17}
A.~N. Heays, A.~D. Bosman and E.~F. van Dishoeck, \emph{Astron. AstroPhys.},
  2017, \textbf{602}, A105\relax
\mciteBstWouldAddEndPuncttrue
\mciteSetBstMidEndSepPunct{\mcitedefaultmidpunct}
{\mcitedefaultendpunct}{\mcitedefaultseppunct}\relax
\EndOfBibitem
\bibitem[Hrodmarsson and van Dishoeck(2023)]{Hrodmarsson-vanDishoeck:23}
H.~R. Hrodmarsson and E.~F. van Dishoeck, \emph{Astron. Astrophys.}, 2023,
  \textbf{675}, A25\relax
\mciteBstWouldAddEndPuncttrue
\mciteSetBstMidEndSepPunct{\mcitedefaultmidpunct}
{\mcitedefaultendpunct}{\mcitedefaultseppunct}\relax
\EndOfBibitem
\bibitem[Yarkony(1995)]{Yarkony:95}
D.~R. Yarkony, \emph{Modern electronic structure theory}, World Scientific,
  Singapore, 1995, p. 642\relax
\mciteBstWouldAddEndPuncttrue
\mciteSetBstMidEndSepPunct{\mcitedefaultmidpunct}
{\mcitedefaultendpunct}{\mcitedefaultseppunct}\relax
\EndOfBibitem
\bibitem[Yarkony(2004)]{Yarkony-cap2-conicalbook:04}
D.~R. Yarkony, \emph{Conical Intersections: electronic structure, dynamics and
  spectroscopy}, Advanced series in Physical Chemistry, World Scientific
  Publishing Co., 2004, p. 175\relax
\mciteBstWouldAddEndPuncttrue
\mciteSetBstMidEndSepPunct{\mcitedefaultmidpunct}
{\mcitedefaultendpunct}{\mcitedefaultseppunct}\relax
\EndOfBibitem
\bibitem[Smith(1969)]{Smith:69}
F.~T. Smith, \emph{Phys. Rev.}, 1969, \textbf{179}, 111\relax
\mciteBstWouldAddEndPuncttrue
\mciteSetBstMidEndSepPunct{\mcitedefaultmidpunct}
{\mcitedefaultendpunct}{\mcitedefaultseppunct}\relax
\EndOfBibitem
\bibitem[Ruedenberg and Atchity(1993)]{Ruedenberg-Atchity:93}
K.~Ruedenberg and G.~J. Atchity, \emph{J. Chem. Phys.}, 1993, \textbf{99},
  3799\relax
\mciteBstWouldAddEndPuncttrue
\mciteSetBstMidEndSepPunct{\mcitedefaultmidpunct}
{\mcitedefaultendpunct}{\mcitedefaultseppunct}\relax
\EndOfBibitem
\bibitem[Mead and Truhlar(1982)]{Mead-Truhlar:82}
C.~A. Mead and D.~G. Truhlar, \emph{J. Chem. Phys.}, 1982, \textbf{77},
  6090\relax
\mciteBstWouldAddEndPuncttrue
\mciteSetBstMidEndSepPunct{\mcitedefaultmidpunct}
{\mcitedefaultendpunct}{\mcitedefaultseppunct}\relax
\EndOfBibitem
\bibitem[K\"oppel(2004)]{Koppel-cap4-conicalbook:04}
H.~K\"oppel, \emph{Conical Intersections: electronic structure, dynamics and
  spectroscopy}, Advanced series in Physical Chemistry, World Scientific
  Publishing Co., 2004, p. 175\relax
\mciteBstWouldAddEndPuncttrue
\mciteSetBstMidEndSepPunct{\mcitedefaultmidpunct}
{\mcitedefaultendpunct}{\mcitedefaultseppunct}\relax
\EndOfBibitem
\bibitem[Thiel and K\"oppel(1999)]{Thiel-Koppel:99}
A.~Thiel and H.~K\"oppel, \emph{J. Chem. Phys.}, 1999, \textbf{110}, 9371\relax
\mciteBstWouldAddEndPuncttrue
\mciteSetBstMidEndSepPunct{\mcitedefaultmidpunct}
{\mcitedefaultendpunct}{\mcitedefaultseppunct}\relax
\EndOfBibitem
\bibitem[Pacher \emph{et~al.}(1991)Pacher, K\"oppel, and
  Cederbaum]{Pacher-etal:91}
T.~Pacher, H.~K\"oppel and L.~S. Cederbaum, \emph{J. Chem. Phys.}, 1991,
  \textbf{95}, 6668\relax
\mciteBstWouldAddEndPuncttrue
\mciteSetBstMidEndSepPunct{\mcitedefaultmidpunct}
{\mcitedefaultendpunct}{\mcitedefaultseppunct}\relax
\EndOfBibitem
\bibitem[Pacher \emph{et~al.}(1993)Pacher, Cederbaum, and
  K\"oppel]{Pacher-etal:93}
T.~Pacher, L.~S. Cederbaum and H.~K\"oppel, \emph{Adv. Chem. Phys.}, 1993,
  \textbf{84}, 293\relax
\mciteBstWouldAddEndPuncttrue
\mciteSetBstMidEndSepPunct{\mcitedefaultmidpunct}
{\mcitedefaultendpunct}{\mcitedefaultseppunct}\relax
\EndOfBibitem
\bibitem[Papas \emph{et~al.}(2008)Papas, Schuurman, and Yarkony]{Papas-etal:08}
B.~N. Papas, M.~S. Schuurman and D.~R. Yarkony, \emph{J. Chem. Phys.}, 2008,
  \textbf{129}, 124104\relax
\mciteBstWouldAddEndPuncttrue
\mciteSetBstMidEndSepPunct{\mcitedefaultmidpunct}
{\mcitedefaultendpunct}{\mcitedefaultseppunct}\relax
\EndOfBibitem
\bibitem[Ghosh \emph{et~al.}(2017)Ghosh, Mukherjee, Mukherjee, Mandal, Sharma,
  {Pinaki Chaudhury}, and Adhikari]{Ghosh-etal:17}
S.~Ghosh, S.~Mukherjee, B.~Mukherjee, S.~Mandal, R.~Sharma, {Pinaki Chaudhury}
  and S.~Adhikari, \emph{J. Chem. Phys.}, 2017, \textbf{147}, 074105\relax
\mciteBstWouldAddEndPuncttrue
\mciteSetBstMidEndSepPunct{\mcitedefaultmidpunct}
{\mcitedefaultendpunct}{\mcitedefaultseppunct}\relax
\EndOfBibitem
\bibitem[Mukherjee \emph{et~al.}(2021)Mukherjee, Ravi, Naskar, Sardar, and
  Adhikari]{Mukherjee-etal:21}
S.~Mukherjee, .~Ravi, K.~Naskar, S.~Sardar and S.~Adhikari, \emph{J. Chem.
  Phys.}, 2021, \textbf{154}, 094306\relax
\mciteBstWouldAddEndPuncttrue
\mciteSetBstMidEndSepPunct{\mcitedefaultmidpunct}
{\mcitedefaultendpunct}{\mcitedefaultseppunct}\relax
\EndOfBibitem
\bibitem[Hazra \emph{et~al.}(2022)Hazra, Mukherjee, Ravi, Sardar, and
  Adhikari]{Hazra-etal:22}
S.~Hazra, S.~Mukherjee, S.~Ravi, S.~Sardar and S.~Adhikari,
  \emph{Chem.Phys.Chem.}, 2022, \textbf{23}, e202200482\relax
\mciteBstWouldAddEndPuncttrue
\mciteSetBstMidEndSepPunct{\mcitedefaultmidpunct}
{\mcitedefaultendpunct}{\mcitedefaultseppunct}\relax
\EndOfBibitem
\bibitem[Guan \emph{et~al.}(2019)Guan, Guo, and Yarkony]{10.1063/1.5099106}
Y.~Guan, H.~Guo and D.~R. Yarkony, \emph{The Journal of Chemical Physics},
  2019, \textbf{150}, 214101\relax
\mciteBstWouldAddEndPuncttrue
\mciteSetBstMidEndSepPunct{\mcitedefaultmidpunct}
{\mcitedefaultendpunct}{\mcitedefaultseppunct}\relax
\EndOfBibitem
\bibitem[Guan \emph{et~al.}(2019)Guan, Zhang, Guo, and Yarkony]{C8CP06598E}
Y.~Guan, D.~H. Zhang, H.~Guo and D.~R. Yarkony, \emph{Phys. Chem. Chem. Phys.},
  2019, \textbf{21}, 14205--14213\relax
\mciteBstWouldAddEndPuncttrue
\mciteSetBstMidEndSepPunct{\mcitedefaultmidpunct}
{\mcitedefaultendpunct}{\mcitedefaultseppunct}\relax
\EndOfBibitem
\bibitem[Guan \emph{et~al.}(2021)Guan, Xie, Yarkony, and Guo]{D1CP03008F}
Y.~Guan, C.~Xie, D.~R. Yarkony and H.~Guo, \emph{Phys. Chem. Chem. Phys.},
  2021, \textbf{23}, 24962--24983\relax
\mciteBstWouldAddEndPuncttrue
\mciteSetBstMidEndSepPunct{\mcitedefaultmidpunct}
{\mcitedefaultendpunct}{\mcitedefaultseppunct}\relax
\EndOfBibitem
\bibitem[Williams and Eisfeld(2018)]{williamsNeuralNetworkDiabatization2018}
D.~M.~G. Williams and W.~Eisfeld, \emph{J. Chem. Phys.}, 2018, \textbf{149},
  204106\relax
\mciteBstWouldAddEndPuncttrue
\mciteSetBstMidEndSepPunct{\mcitedefaultmidpunct}
{\mcitedefaultendpunct}{\mcitedefaultseppunct}\relax
\EndOfBibitem
\bibitem[Xie \emph{et~al.}(2018)Xie, Zhu, Yarkony, and Guo]{Xie-etal:18}
C.~Xie, X.~Zhu, D.~R. Yarkony and H.~Guo, \emph{The Journal of Chemical
  Physics}, 2018, \textbf{149}, 144107\relax
\mciteBstWouldAddEndPuncttrue
\mciteSetBstMidEndSepPunct{\mcitedefaultmidpunct}
{\mcitedefaultendpunct}{\mcitedefaultseppunct}\relax
\EndOfBibitem
\bibitem[Li \emph{et~al.}(2024)Li, Liang, Niu, He, Xing, and
  Zhang]{Li-etal_SiH2p:24}
W.~Li, Y.~Liang, X.~Niu, D.~He, W.~Xing and Y.~Zhang, \emph{The Journal of
  Chemical Physics}, 2024, \textbf{161}, 044310\relax
\mciteBstWouldAddEndPuncttrue
\mciteSetBstMidEndSepPunct{\mcitedefaultmidpunct}
{\mcitedefaultendpunct}{\mcitedefaultseppunct}\relax
\EndOfBibitem
\bibitem[Li \emph{et~al.}(2024)Li, Dong, Niu, Wang, and Zhang]{Li-etal_CH2p:24}
W.~Li, B.~Dong, X.~Niu, M.~Wang and Y.~Zhang, \emph{The Journal of Chemical
  Physics}, 2024,  074302\relax
\mciteBstWouldAddEndPuncttrue
\mciteSetBstMidEndSepPunct{\mcitedefaultmidpunct}
{\mcitedefaultendpunct}{\mcitedefaultseppunct}\relax
\EndOfBibitem
\bibitem[Shu and Truhlar(2020)]{Shu-Truhlar:20}
Y.~Shu and D.~G. Truhlar, \emph{J. Chem. The. Comp.}, 2020, \textbf{16},
  6456\relax
\mciteBstWouldAddEndPuncttrue
\mciteSetBstMidEndSepPunct{\mcitedefaultmidpunct}
{\mcitedefaultendpunct}{\mcitedefaultseppunct}\relax
\EndOfBibitem
\bibitem[Shu \emph{et~al.}(2024)Shu, Akher, Guo, and
  Truhlar]{shuParametricallyManagedActivation2024}
Y.~Shu, F.~B. Akher, H.~Guo and D.~G. Truhlar, \emph{J. Phys. Chem. A}, 2024,
  \textbf{128}, 1207--1217\relax
\mciteBstWouldAddEndPuncttrue
\mciteSetBstMidEndSepPunct{\mcitedefaultmidpunct}
{\mcitedefaultendpunct}{\mcitedefaultseppunct}\relax
\EndOfBibitem
\bibitem[{del Mazo-Sevillano} \emph{et~al.}(2025){del Mazo-Sevillano}, Aguado,
  Lique, {Jara-Toro}, and Roncero]{delMazo25_etal_CHp}
P.~{del Mazo-Sevillano}, A.~Aguado, F.~Lique, R.~A. {Jara-Toro} and O.~Roncero,
  \emph{Phys. Chem. Chem. Phys.}, 2025, \textbf{27}, 15775--15786\relax
\mciteBstWouldAddEndPuncttrue
\mciteSetBstMidEndSepPunct{\mcitedefaultmidpunct}
{\mcitedefaultendpunct}{\mcitedefaultseppunct}\relax
\EndOfBibitem
\bibitem[se\ n \emph{et~al.}(2024)se\ n, von Lilienfeld, and
  cek]{Srsen-etal:24}
S.~S. se\ n, O.~A. von Lilienfeld and P.~S. cek, \emph{PCCP}, 2024,
  \textbf{26}, 4306\relax
\mciteBstWouldAddEndPuncttrue
\mciteSetBstMidEndSepPunct{\mcitedefaultmidpunct}
{\mcitedefaultendpunct}{\mcitedefaultseppunct}\relax
\EndOfBibitem
\bibitem[Jensen \emph{et~al.}(1995)Jensen, Brumm, Kraemer, and
  Bunker]{Jensen-etal:95}
P.~Jensen, M.~Brumm, W.~P. Kraemer and P.~R. Bunker, \emph{J. Mol. Spectros.},
  1995, \textbf{172}, 194\relax
\mciteBstWouldAddEndPuncttrue
\mciteSetBstMidEndSepPunct{\mcitedefaultmidpunct}
{\mcitedefaultendpunct}{\mcitedefaultseppunct}\relax
\EndOfBibitem
\bibitem[{Black} and {Dalgarno}(1976)]{Black-Dalgarno:76}
J.~H. {Black} and A.~{Dalgarno}, \emph{AstroPhys. J.}, 1976, \textbf{203},
  132--142\relax
\mciteBstWouldAddEndPuncttrue
\mciteSetBstMidEndSepPunct{\mcitedefaultmidpunct}
{\mcitedefaultendpunct}{\mcitedefaultseppunct}\relax
\EndOfBibitem
\bibitem[Smith(1992)]{Smith:92}
D.~Smith, \emph{Chem. Rev.}, 1992, \textbf{92}, 1473\relax
\mciteBstWouldAddEndPuncttrue
\mciteSetBstMidEndSepPunct{\mcitedefaultmidpunct}
{\mcitedefaultendpunct}{\mcitedefaultseppunct}\relax
\EndOfBibitem
\bibitem[Douglas and Herzberg(1941)]{Douglas-Herzberg:41}
A.~E. Douglas and G.~Herzberg, \emph{AstroPhys. J.}, 1941, \textbf{94},
  381\relax
\mciteBstWouldAddEndPuncttrue
\mciteSetBstMidEndSepPunct{\mcitedefaultmidpunct}
{\mcitedefaultendpunct}{\mcitedefaultseppunct}\relax
\EndOfBibitem
\bibitem[Bern{\'e} \emph{et~al.}(2023)Bern{\'e}, Martin-Drumel, Schroetter, and
  {\it et al.}]{Berne-etal:23}
O.~Bern{\'e}, M.-A. Martin-Drumel, I.~Schroetter and {\it et al.},
  \emph{Nature}, 2023, \textbf{621}, 56\relax
\mciteBstWouldAddEndPuncttrue
\mciteSetBstMidEndSepPunct{\mcitedefaultmidpunct}
{\mcitedefaultendpunct}{\mcitedefaultseppunct}\relax
\EndOfBibitem
\bibitem[Saxon \emph{et~al.}(1980)Saxon, Kirby, and Liu]{Saxon-etal:80}
R.~P. Saxon, K.~Kirby and B.~Liu, \emph{J. Chem. Phys.}, 1980, \textbf{73},
  1873\relax
\mciteBstWouldAddEndPuncttrue
\mciteSetBstMidEndSepPunct{\mcitedefaultmidpunct}
{\mcitedefaultendpunct}{\mcitedefaultseppunct}\relax
\EndOfBibitem
\bibitem[Kirby \emph{et~al.}(1980)Kirby, Roberge, Saxon, and
  Liu]{Kirby-etal:80}
K.~Kirby, W.~G. Roberge, R.~P. Saxon and B.~Liu, \emph{AstroPhys. J.}, 1980,
  855\relax
\mciteBstWouldAddEndPuncttrue
\mciteSetBstMidEndSepPunct{\mcitedefaultmidpunct}
{\mcitedefaultendpunct}{\mcitedefaultseppunct}\relax
\EndOfBibitem
\bibitem[del Mazo-Sevillano \emph{et~al.}(2024)del Mazo-Sevillano, Aguado,
  Goicoechea, and Roncero]{delMazo-Sevillano-etal:24}
P.~del Mazo-Sevillano, A.~Aguado, J.~R. Goicoechea and O.~Roncero, \emph{J.
  Chem. Phys.}, 2024, \textbf{160}, 184307\relax
\mciteBstWouldAddEndPuncttrue
\mciteSetBstMidEndSepPunct{\mcitedefaultmidpunct}
{\mcitedefaultendpunct}{\mcitedefaultseppunct}\relax
\EndOfBibitem
\bibitem[Theodorakopoulos and Petsalakis(1991)]{Theodorakopoulos-Petsalakis:91}
G.~Theodorakopoulos and I.~D. Petsalakis, \emph{J. Mol. Structure (Theochem)},
  1991, \textbf{230}, 205\relax
\mciteBstWouldAddEndPuncttrue
\mciteSetBstMidEndSepPunct{\mcitedefaultmidpunct}
{\mcitedefaultendpunct}{\mcitedefaultseppunct}\relax
\EndOfBibitem
\bibitem[Jiang and Guo(2013)]{Jiang-Guo:13}
B.~Jiang and H.~Guo, \emph{J. Chem. Phys.}, 2013, \textbf{139}, 054112\relax
\mciteBstWouldAddEndPuncttrue
\mciteSetBstMidEndSepPunct{\mcitedefaultmidpunct}
{\mcitedefaultendpunct}{\mcitedefaultseppunct}\relax
\EndOfBibitem
\bibitem[Li \emph{et~al.}(2020)Li, Varga, Truhlar, and Guo]{JLi2020}
J.~Li, Z.~Varga, D.~G. Truhlar and H.~Guo, \emph{Journal of Chemical Theory and
  Computation}, 2020, \textbf{16}, 4822--4832\relax
\mciteBstWouldAddEndPuncttrue
\mciteSetBstMidEndSepPunct{\mcitedefaultmidpunct}
{\mcitedefaultendpunct}{\mcitedefaultseppunct}\relax
\EndOfBibitem
\bibitem[del Mazo-Sevillano \emph{et~al.}(2024)del Mazo-Sevillano,
  F{\'e}lix-Gonz{\'a}lez, Aguado, Sanz-Sanz, Kwon, and
  Roncero]{delMazo-Sevillano-etal:24a}
P.~del Mazo-Sevillano, D.~F{\'e}lix-Gonz{\'a}lez, A.~Aguado, C.~Sanz-Sanz,
  D.~Kwon and O.~Roncero, \emph{Mol. Phys.}, 2024,  e2183071\relax
\mciteBstWouldAddEndPuncttrue
\mciteSetBstMidEndSepPunct{\mcitedefaultmidpunct}
{\mcitedefaultendpunct}{\mcitedefaultseppunct}\relax
\EndOfBibitem
\bibitem[Houston \emph{et~al.}(2024)Houston, Qu, Yu, Pandey, Conte, Nandi, and
  Bowman]{Houston2024}
P.~L. Houston, C.~Qu, Q.~Yu, P.~Pandey, R.~Conte, A.~Nandi and J.~M. Bowman,
  \emph{Journal of Chemical Theory and Computation}, 2024, \textbf{20},
  3008--3018\relax
\mciteBstWouldAddEndPuncttrue
\mciteSetBstMidEndSepPunct{\mcitedefaultmidpunct}
{\mcitedefaultendpunct}{\mcitedefaultseppunct}\relax
\EndOfBibitem
\bibitem[Neese(2025)]{RN269}
F.~Neese, \emph{WIRES Comput. Molec. Sci.}, 2025, \textbf{15}, e70019\relax
\mciteBstWouldAddEndPuncttrue
\mciteSetBstMidEndSepPunct{\mcitedefaultmidpunct}
{\mcitedefaultendpunct}{\mcitedefaultseppunct}\relax
\EndOfBibitem
\bibitem[Kollmar \emph{et~al.}(2019)Kollmar, Sivalingam, Helmich-Paris, Angeli,
  and Neese]{RN176}
C.~Kollmar, K.~Sivalingam, B.~Helmich-Paris, C.~Angeli and F.~Neese, \emph{J.
  Comput. Chem.}, 2019, \textbf{40}, 1463--1470\relax
\mciteBstWouldAddEndPuncttrue
\mciteSetBstMidEndSepPunct{\mcitedefaultmidpunct}
{\mcitedefaultendpunct}{\mcitedefaultseppunct}\relax
\EndOfBibitem
\bibitem[Neese(2022)]{RN232}
F.~Neese, \emph{J. Comp. Chem.}, 2022, \textbf{44}, 381\relax
\mciteBstWouldAddEndPuncttrue
\mciteSetBstMidEndSepPunct{\mcitedefaultmidpunct}
{\mcitedefaultendpunct}{\mcitedefaultseppunct}\relax
\EndOfBibitem
\bibitem[Ugandi and Roemelt(2023)]{RN243}
M.~Ugandi and M.~Roemelt, \emph{Int. J. Quantum Chem.}, 2023, \textbf{123},
  e27045\relax
\mciteBstWouldAddEndPuncttrue
\mciteSetBstMidEndSepPunct{\mcitedefaultmidpunct}
{\mcitedefaultendpunct}{\mcitedefaultseppunct}\relax
\EndOfBibitem
\bibitem[Sanz-Sanz \emph{et~al.}(2015)Sanz-Sanz, Aguado, Roncero, and
  Naumkin]{Sanz-Sanz-etal:15}
C.~Sanz-Sanz, A.~Aguado, O.~Roncero and F.~Naumkin, \emph{J. Chem. Phys.},
  2015, \textbf{143}, 234303\relax
\mciteBstWouldAddEndPuncttrue
\mciteSetBstMidEndSepPunct{\mcitedefaultmidpunct}
{\mcitedefaultendpunct}{\mcitedefaultseppunct}\relax
\EndOfBibitem
\bibitem[Roncero and del Mazo-Sevillano(2025)]{Roncero-delMazo-Sevillano:25}
O.~Roncero and P.~del Mazo-Sevillano, \emph{Comp. Phys. Comm.}, 2025,
  109471\relax
\mciteBstWouldAddEndPuncttrue
\mciteSetBstMidEndSepPunct{\mcitedefaultmidpunct}
{\mcitedefaultendpunct}{\mcitedefaultseppunct}\relax
\EndOfBibitem
\bibitem[Paniagua \emph{et~al.}(1999)Paniagua, Aguado, Lara, and
  Roncero]{Paniagua-etal:99}
M.~Paniagua, A.~Aguado, M.~Lara and O.~Roncero, \emph{J. Chem. Phys.}, 1999,
  \textbf{111}, 6712\relax
\mciteBstWouldAddEndPuncttrue
\mciteSetBstMidEndSepPunct{\mcitedefaultmidpunct}
{\mcitedefaultendpunct}{\mcitedefaultseppunct}\relax
\EndOfBibitem
\bibitem[Aguado \emph{et~al.}(2003)Aguado, Paniagua, Sanz-Sanz, and
  Roncero]{Aguado-etal:03}
A.~Aguado, M.~Paniagua, C.~Sanz-Sanz and O.~Roncero, \emph{J. Chem. Phys.},
  2003, \textbf{119}, 10088\relax
\mciteBstWouldAddEndPuncttrue
\mciteSetBstMidEndSepPunct{\mcitedefaultmidpunct}
{\mcitedefaultendpunct}{\mcitedefaultseppunct}\relax
\EndOfBibitem
\bibitem[Chenel \emph{et~al.}(2016)Chenel, Roncero, Aguado, Ag\'undez, and
  Cernicharo]{Chenel-etal:16}
A.~Chenel, O.~Roncero, A.~Aguado, M.~Ag\'undez and J.~Cernicharo, \emph{J.
  Chem. Phys.}, 2016, \textbf{144}, 144306\relax
\mciteBstWouldAddEndPuncttrue
\mciteSetBstMidEndSepPunct{\mcitedefaultmidpunct}
{\mcitedefaultendpunct}{\mcitedefaultseppunct}\relax
\EndOfBibitem
\bibitem[Aguado \emph{et~al.}(2017)Aguado, Roncero, Zanchet, Ag{\'u}ndez, and
  Cernicharo]{Aguado-etal:17}
A.~Aguado, O.~Roncero, A.~Zanchet, M.~Ag{\'u}ndez and J.~Cernicharo,
  \emph{Astrophys. J.}, 2017, \textbf{838}, 33\relax
\mciteBstWouldAddEndPuncttrue
\mciteSetBstMidEndSepPunct{\mcitedefaultmidpunct}
{\mcitedefaultendpunct}{\mcitedefaultseppunct}\relax
\EndOfBibitem
\bibitem[G{\'o}mez-Carrasco and Roncero(2006)]{Gomez-Carrasco-Roncero:06}
S.~G{\'o}mez-Carrasco and O.~Roncero, \emph{J. Chem. Phys.}, 2006,
  \textbf{125}, 054102\relax
\mciteBstWouldAddEndPuncttrue
\mciteSetBstMidEndSepPunct{\mcitedefaultmidpunct}
{\mcitedefaultendpunct}{\mcitedefaultseppunct}\relax
\EndOfBibitem
\bibitem[Draine(1978)]{Draine:78}
Draine, \emph{AstroPhysical J. Suppl. Ser.}, 1978, \textbf{36}, 595\relax
\mciteBstWouldAddEndPuncttrue
\mciteSetBstMidEndSepPunct{\mcitedefaultmidpunct}
{\mcitedefaultendpunct}{\mcitedefaultseppunct}\relax
\EndOfBibitem
\end{mcitethebibliography}
\providecommand*{\mcitethebibliography}{\thebibliography}
\csname @ifundefined\endcsname{endmcitethebibliography}
{\let\endmcitethebibliography\endthebibliography}{}

\end{multicols}

\end{document}